\documentclass{article}
\usepackage{amsmath,graphicx,mlspconf}
\usepackage{xcolor}
\usepackage{hyperref}
\usepackage{amsmath,amssymb,amsfonts}
\usepackage{algorithm}
\usepackage{algorithmicx}
\usepackage{algpseudocode}  
\usepackage{comment}
\usepackage{textcomp}
\usepackage{xcolor}
\usepackage{lscape,lipsum}
\usepackage{threeparttable}
\usepackage{longtable}
\usepackage{multirow}
\usepackage{makecell}
\usepackage{etoolbox}
\usepackage{pgfplots} 
\usepackage{tikz}
\usepackage{graphicx, adjustbox}
\usepackage{subfig}
\usepackage{mathtools}
\usepackage{array}
\usepackage{booktabs}
\usepackage[usestackEOL]{stackengine}
\usepackage{xfrac}
\usepackage{threeparttable}
\usetikzlibrary{shapes.geometric}
\usetikzlibrary{3d}
\usepackage{url}
\usepackage{mathtools}
\usepackage{siunitx} 
\title{Signal Prediction for Loss Mitigation in Tactile Internet: \\A Leader-Follower Game-Theoretic Approach}
\name{Mohammad Ali Vahedifar, and Qi Zhang \thanks{
\\Authors' e-mails: \{av, qz\}@ece.au.dk.\\
Link to the Code: \href{https://github.com/Ali-Vahedifar/Leader-Follower-LeFo.git}{GitHub Repository}.}}
\address{DIGIT and Department of Electrical and Computer Engineering, Aarhus University, Denmark.}

\begin{document}

\maketitle

\begin{abstract}
Tactile Internet (TI) requires achieving ultra-low latency and highly reliable packet delivery for haptic signals. In the presence of packet loss and delay, the signal prediction method provides a viable solution for recovering the missing signals. To this end, we introduce the Leader-Follower (LeFo) approach based on a cooperative Stackelberg game, which enables both users and robots to learn and predict actions. With accurate prediction, the teleoperation system can safely relax its strict delay requirements.  Our method achieves high prediction accuracy, ranging from 80.62\% to 95.03\% for remote robot signals at the Human ($\mathcal{H}$) side and from 70.44\% to 89.77\% for human operation signals at the remote Robot ($\mathcal{R}$) side. We also establish an upper bound for maximum signal loss using Taylor Expansion, ensuring robustness.  
\end{abstract}
\section{Introduction}
Tactile Internet (TI) is defined as ``a network, or a network of networks, for remotely accessing, perceiving, manipulating, or controlling real and virtual objects or processes in perceived real-time"~\cite {Promwongsa}. TI enables the communication of touch, allowing for immersive Human-to-Machine/Robot (H2M/R) interactions and manipulations over large physical distances~\cite{Antonakoglou}. To achieve transparent and immersive teleoperation and telemanipulation, TI demands haptic packet transmission with ultra-low latency (even as low as 1 $ms$) and high reliability~\cite{zhang20185genabledtactilerobotic}. Meeting these stringent requirements poses significant challenges in communication systems.

In this paper, we address the challenge of maintaining reliable performance in TI systems by recovering from a communication outage up to 1 sample of latency.  We propose a novel Game Theory (GT)-based signal prediction model for TI, the \emph{Leader-Follower} (LeFo) framework, as the interaction between a human and a robot can be modeled as a two-agent Stackelberg game. We aim to utilize bidirectional predictive modeling, the LeFo, to overcome packet loss and relax latency constraints for TI services, thereby enabling human and robotic agents to predict signals. On the human side, the system predicts the robot’s subsequent force, velocity, and position at sequence step $n+1$. Conversely, on the robot side, it forecasts the human’s forthcoming force, velocity, and position at sequence step $n+1$. Our approach formulates a \textit{MiniMax} optimization problem with iterative learning loops to enhance real-time responsiveness. Furthermore, we derive an upper bound on the training loss using a Taylor expansion for missing signal estimation in TI, providing a theoretical guarantee for model robustness.
 \begin{figure*}[tbp]
    \centering
    \includegraphics[width=\textwidth]{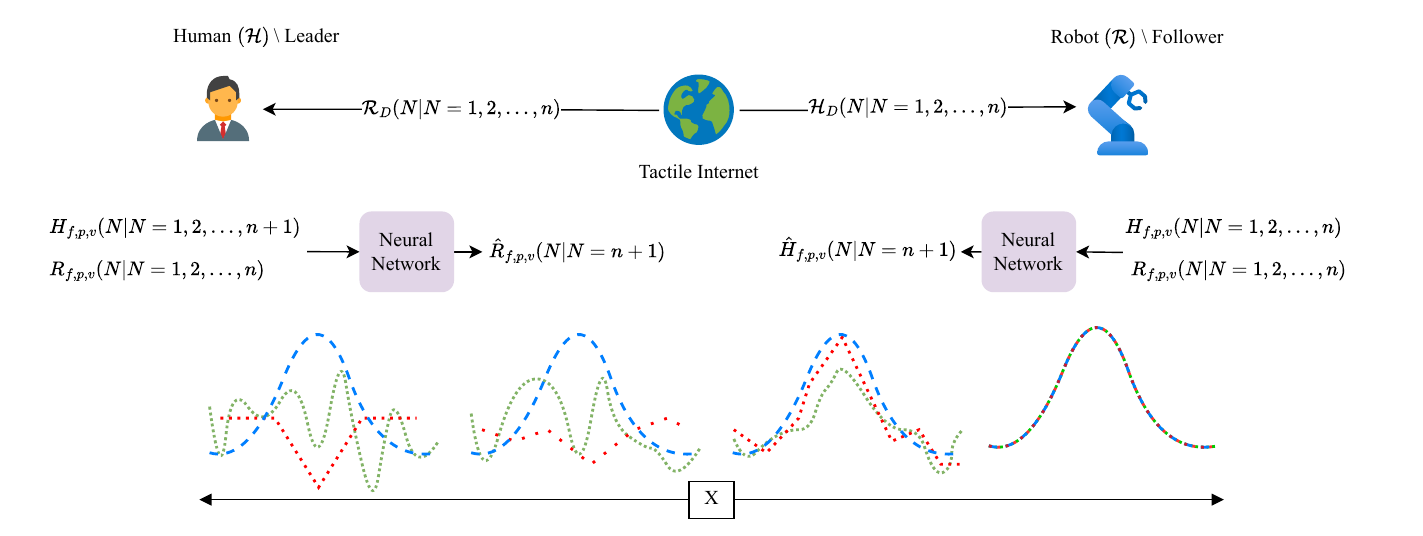}
    \caption{This diagram illustrates the conceptual framework of the LeFo method. Each Neural Network (NN) processes past human and robot data samples, gathered from the start up to the current prediction point ($N=1,2,..., n$), to predict the very next sample. Specifically, on the robot side at time $N=n$, it predicts the human's next sample ($\hat{H}(n+1)$)  and, on the human side after receiving ($R(n)$), the system predicts the robot's next sample ($\hat{R}(n+1)$).  The terms $\mathcal{H}_D$ and $\mathcal{R}_D$ represent the ground truth for evaluation. The robot feedback in $\mathcal{R}(N)$ is the response to the human signal in $\mathcal{H}(N)$.}
    \label{fig: Conceptual}
\end{figure*} 

GT has been extensively applied in wireless communication and edge computing to optimize critical aspects, such as cognitive radio~\cite{Wang10.1109/COMST.2016.2539923}, caching strategies~\cite{chen2017green},  and decision-making~\cite{BENSASSIgametheory}. These applications are particularly prominent in vehicular networks~\cite{Jiang9964256},  radio resource slicing~\cite {kokkinis2025deepreinforcementlearningbasedvideohaptic}, and edge computing~\cite{Hou9681620}. While existing studies have demonstrated the efficacy of GT approaches in wireless communications, no work has leveraged the GT approach for haptic signal prediction to enhance semi-autonomous actions for human-in-the-loop interaction. Our work addresses this gap by introducing a novel Stackelberg game-based framework explicitly tailored to the unique requirements of TI systems.

\section{Leader-Follower methodology}
In the Stackelberg game, the human possesses prior knowledge of system dynamics, while the robot optimally adapts its response based on the human's actions. Both agents optimize their respective utility functions through strategic interaction, where the Leader's decisions influence the Follower's responses, and the Follower reacts to maximize its utility as shown in Fig.~\ref{fig: Conceptual}. The utility function serves as a mathematical representation of an agent's preferences over a set of possible outcomes. Formally, it assigns a real-valued measure to each outcome within the strategy space, enabling the analysis of optimal decision-making under both non-cooperative and cooperative interaction paradigms.

\subsection{Leader-Follower game background}

In the Leader-Follower game, we have \( S_L \), which represents the Leader's signal or alternatively strategy. Similarly, \( S_F \) denotes the Follower's response signal or strategy. The Follower optimizes its feedback by solving:
\begin{equation}
     S^*_F(S_L) = \arg\max_{s_F} U_F(S_F, S_L),
\end{equation}
where \( U_F \) is the Follower’s utility function.  The Leader optimizes its utility function  \( U_L\leftarrow U_L\big(S^*_F(S_L), S_L\big) \) while considering the Follower’s response, ensuring that \( U_L \) is maximized based on the Follower’s optimal reaction. The Leader’s optimal strategy is obtained by solving:
\begin{equation}
    S^*_L = \arg\max_{s_L} U_L\big(S^*_F(S_L), S_L\big). 
\end{equation}
This ensures that the leader’s decision leads to the most effective interaction. At equilibrium, the system achieves a closed-loop structure:~\begin{equation}
    S^*_F = S^*_F(S^*_L).
\end{equation}
\subsection{Leader-Follower MiniMax game mathematical formulation}
Motivated by the Leader-Follower game background. Let us indicate Leader as Human ($\mathcal{H}$) and Follower as Robot ($\mathcal{R}$) in TI. The prediction of the next samples for both sides is:
\begin{equation}
\hat{S}_{A}(n+1) =  \sum_{i=1}^N \Omega_i S_{A}(n+1-i) + \sum_{j=1}^N \lambda_j \epsilon_{n+1-j},~A \hspace{-1mm}\in\hspace{-1mm}{\{\mathcal{R,\hspace{-1mm} H}\}}.
\end{equation}

\noindent Here $\Omega_i$ is the $i$-th auto-regressive coefficient, which determines the influence of the $i$-th past human signal on the current prediction. $\lambda_j$ is the $j$-th moving average coefficient, which determines the influence of the $j$-th past error term on the current prediction. $\epsilon_{n+1-j}$ is the error term at time $n+1-j$. For future time steps (i.e., when $n+1-j > n$), these error terms are generally assumed to be zero for point forecasts.

The loss function for the ($\mathcal{R}$) side is to minimize the Kullback–Leibler ($\textit{KL}$) divergence between the actual ($S_\mathcal{H}(n+1)$) and predicted ($\hat{S}_\mathcal{H}(n+1)$) signals of $\mathcal{H}$.
\begin{align}
     U_\mathcal{R}\big(S_\mathcal{H}(n),S_\mathcal{R}(n)\big)= \textit{KL}_\mathcal{R}\big(S_\mathcal{H}(n+1), \hat{S}_\mathcal{H}(n+1)\big).\\
     \hat{\mathcal{H}}_{f,p,v}(n+1)\gets \arg\min U_\mathcal{R}\big(S_\mathcal{H}(n),S_\mathcal{R}(n)\big).
\end{align}
For the next step, the leader $\mathcal{H}$ receives $S_\mathcal{R}(n)$. Then $\mathcal{H}$ calculates its object utility function, which we defined as Mutual Information, $I$, between actual response feedback $S_\mathcal{R}(n+1)$ and predicted feedback $\hat{S}_\mathcal{R}(n+1)$ of $\mathcal{R}$.
\begin{align}
&\hspace{-2mm} U_\mathcal{H}\big(S_\mathcal{H}(n+1),S_\mathcal{R}(n)\big) =  I_\mathcal{H}\big(S_\mathcal{R}(n+1); \hat{S}_\mathcal{R}(n+1)\big).\hspace{-1mm}\\
&\hspace{-2mm} \hat{\mathcal{R}}_{f,p,v}(n+1) \gets \arg\max U_\mathcal{H}\big(S_\mathcal{H}(n+1),S_\mathcal{R}(n)\big).
\end{align}

\begin{figure*}[!ht]
\begin{equation}
\begin{split}
\min_\mathcal{R} \max_\mathcal{H} \mathcal{L}(\mathcal{H},\mathcal{R},\theta_n)=\mathbb{E}_{S_\mathcal{H}}\bigg[\max  I_\mathcal{H}\big(S_\mathcal{R}(n+1); \hat{S}_\mathcal{R}(n+1)\big)\bigg] - \mathbb{E}_{S_\mathcal{R}}\bigg[\min \textit{KL}_\mathcal{R}\big(S_\mathcal{H}(n+1), \hat{S}_\mathcal{H}(n+1)\big)\bigg].
\end{split} \label{Eq: MiniMax}
\end{equation}
\end{figure*}

\noindent Given the $S_\mathcal{H}(n)$’s signal, the follower $\mathcal{R}$ determines its best response. Mathematically, the interaction is formulated as a MiniMax optimization problem presented in Eq.~\ref{Eq: MiniMax} with $\theta_n$ as a parameter of NN for predicting $S_A(n), A\in\{\mathcal{R,H}\}$. The LeFo pseudo code is shown in Algorithm~\ref{alg: lefo_mi_knn}.

\subsection{Estimation of the mutual information using K-nearest neighbors}
Following the work of~\cite{vahedifar2024informationmodifiedknearestneighbor,  AMIRI20111184}, we estimate the \( I(S_\mathcal{R},\hat{S}_\mathcal{R}) \) using the K-Nearest Neighbors (KNN) method. Here, $K$ is the number of nearest neighbors. We use simplification in notation by removing the subscripts and simply using $S$ and $\hat{S}$. Given a dataset of \( N \) leader-follower signal pairs:
\begin{equation}
z^j_i = ( S_i^j, \hat{S}_i^j),\quad j=f,v,p,\quad i = 1, \dots, N.
\end{equation}
Here, j represents the features: force ($f$), position ($p$), and velocity ($v$).
The samples are drawn from an unknown joint probability distribution \(p_{S_i, \hat{S}_i}(S_i, \hat{S}_i) \). Here, $I$ quantifies the dependency between \(S_i\) and \(\hat{S}_i\) and is given by:
\begin{equation}
I(S_i; \hat{S}_i) = H(S_i) + H(\hat{S}_i) - H(S_i, \hat{S}_i),
\end{equation}
where \( H(S_i) \) and  \( H(\hat{S}_i) \) are the marginal entropies of \( S_\mathcal{R} \) and \( \hat{S}_i \) , and \( H(S_i, \hat{S}_i) \) is the joint entropy. Since the true distribution \( p_{S_i, \hat{S}_i}(S_i, \hat{S}_i) \) is unknown,
$I$ is approximated using a non-parametric estimation technique based on the KNN method. The Chebyshev norm definition:
\begin{equation}
\|z_i - z_u\|_\infty = \max \left( \|x_i - x_u\|, \|y_i - y_u\| \right).
\end{equation}
To measure distances between points in the dataset, the Chebyshev norm is used:
\begin{equation}
\begin{split}
||z_i - \hat{z}_i||_\infty = \max \big( ||S_i^f - \hat{S}_i^f||, ||S_i^v - \hat{S}_i^v||,\\ ||S_i^p - \hat{S}_i^p|| \big), \qquad i = 1, \dots, N.
\end{split}
\end{equation}

\noindent For each point \( z_i^j \), the \( k \)-th nearest neighbor is denoted as:
\begin{equation}\label{k nearest neigbour}
z_i^{j,k} = (S_i^{j,k}, \hat{S}_i^{j,k}).
\end{equation}
The distances between \( z_i^j \) and its \( k \)-th nearest neighbor are given by:
\begin{align}
e^f &= 2||S_i^f - \hat{S}_i^{f,k}||, \label{Eq: ef}\\
e^v &= 2||S_i^v - \hat{S}_i^{v,k}||, \label{Eq: ev}\\
e^p &= 2||S_i^p - \hat{S}_i^{p,k}||.\label{Eq: ep}
\end{align}
\noindent Define neighbor counts:
\begin{align}
n^f: \text{Number of points with}\quad e^f \geq 2||S_i^f - \hat{S}_i^{f,k}||,   \label{Eq: nf}\\
n^v: \text{Number of points with}\quad e^v \geq 2||S_i^v - \hat{S}_i^{v,k}||,  \label{Eq: nv}\\
n^p: \text{Number of points with}\quad e^p \geq 2||S_i^p - \hat{S}_i^{p,k}||. \label{Eq: np}
\end{align}
\noindent The $I$ estimator is:
\begin{align}
I(S_\mathcal{R}; \hat{S}_\mathcal{R}) &=  \psi(N) - \frac{1}{N} \sum_{i=1}^{N} \left( \psi(n^f) + \psi(n^v) + \psi(n^p) \right) \nonumber\\
&+ \psi(K) - \frac{1}{K}, \label{Eq: MI}
\end{align}

\noindent where, \( \psi(n) \) is the digamma function. It is defined as:
\begin{equation}
\psi(n) = \frac{d}{dn} \ln \Gamma(n) = \frac{\Gamma'(n)}{\Gamma(n)},
\end{equation}
which approximates the logarithm of a factorial. In this paper, we use the 11-th nearest neighbor to estimate $I$.

\begin{algorithm}[t]
\caption{Leader-Follower (LeFo) Algorithm}
\label{alg: lefo_mi_knn}
\begin{algorithmic}
\State Initialize $S_\mathcal{H}$ and $S_\mathcal{R}$, set max\_iterations, tolerance
\Function{Utility\_Robot}{$S_\mathcal{H}, \hat{S}_\mathcal{H}$}
\State $\textit{KL}(S_\mathcal{H} || \hat{S}_\mathcal{H}) = \sum S_\mathcal{H} \log \frac{S_\mathcal{H}}{\hat{S}_\mathcal{H}}$
\State \Return $\textit{KL}(S_\mathcal{H}, \hat{S}_\mathcal{H})$ 
\EndFunction
\Function{Utility\_Human}{$S_\mathcal{R}, \hat{S}_\mathcal{R}$}
\For {each sample in $(S^j_i, \hat{S}^j_i)$}
        \State Find the $k$-th nearest neighbor with Eq.~\ref{k nearest neigbour}.
        \State Compute the Chebyshev norms with Eq.~\ref{Eq: ef}, \State~\ref{Eq: ev},~and~\ref{Eq: ep}.
        \State Define neighbor counts with Eq.~\ref{Eq: nf},~\ref{Eq: nv},~and~\ref{Eq: np}.
\EndFor
\State Compute $I$ with Eq.~\ref{Eq: MI}
\State \Return $I(S_\mathcal{R}; \hat{S}_\mathcal{R})$
\EndFunction
\For{iteration = 1 \textbf{to} max\_iterations}
\State $U_\mathcal{R} \gets \arg\min_{S_\mathcal{R}} \textsc{Utility\_Robot}(S_\mathcal{H}, \hat{S}_\mathcal{H})$ 
\State $U_\mathcal{H} \gets \arg\max_{S_\mathcal{H}} \textsc{Utility\_Human}(S_\mathcal{R},\hat{S}_\mathcal{R})$ 
\State Solve MiniMax optimization with Eq.~\ref{Eq: MiniMax}
\If{$\| \mathcal{L}(\mathcal{H},\mathcal{R},\theta_n) - \mathcal{L}^*(\mathcal{H},\mathcal{R},\theta_n)) \| < $ tolerance}
\State \textbf{Break}
\EndIf
\EndFor
\State \Return $\hat{S}_\mathcal{H}, \hat{S}_\mathcal{R}$
\end{algorithmic}
\end{algorithm}
\subsection{Upper bound for loss function in Tactile Internet}
\begin{figure*}[t]
    \centering
    \includegraphics[width=\textwidth]{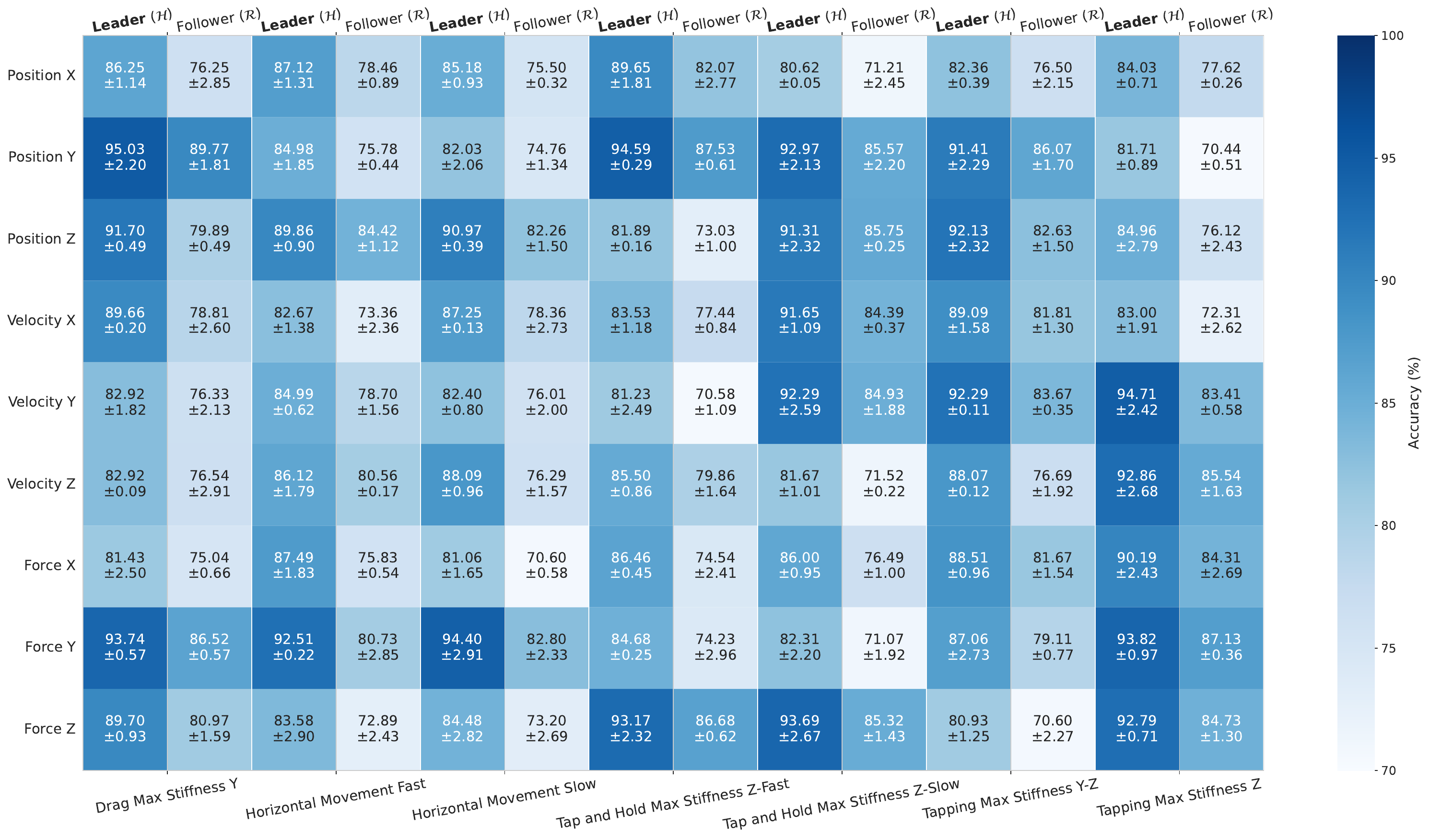}
    \caption{Prediction accuracy and standard deviation for each dataset using parameter set: Max Stiffness = 0.4, Static Friction = 1.0, Dynamic Friction = 0.7, and No Viscosity. Results are averaged over 10 independent runs.}
    \label{fig: Prediction accuracy}
\end{figure*}

Let us define by $\theta_{n-1}$ the parameters of an NN model trained on the previous signal $S_{n-1}$, which are used to obtain the parameters $\theta_n$ by updating the model to learn a new signal $S_n$ with Eq.~\ref{Eq: MiniMax}. The total loss on the predicted signal and the ground truth takes the form:
\begin{equation}
     \mathcal{L}_{n}(\mathcal{H},\mathcal{R},\theta_n) = \frac{1}{|S_n|} \sum_{ S_n} \ell_n(\theta_n; S_n).
\end{equation}
Let us define loss as:
\begin{equation}
    \mathcal{L} \overset{\underset{\mathrm{def}}{}}{=} \mathcal{L}_{n-1}(\mathcal{H},\mathcal{R},\theta_n) - \mathcal{L}_{n-1}(\mathcal{H},\mathcal{R},\theta_{n-1}).
\end{equation}
For simplicity, we remove $\mathcal{H}$ and $\mathcal{R}$ from the notation. In the general case, \(\theta_{n-1}\) and \(\theta_n\) correspond to the optimal parameters obtained after training has been completed for predicting the signal $S_{n-1}$ and $S_n$, respectively. Assuming that these optimal points are close to each other, the Taylor expansion of the loss function \(\mathcal{L}_{n-1}(\theta_n)\) around \(\theta_{n-1}\) up to the second order is given by:~\begin{equation}
\label{Eq:TaylorExpL1Theta2} 
\begin{aligned} 
&\mathcal{L}_{n-1}(\theta_n) \approx \mathcal{L}_{n-1}(\theta_{n-1}) + (\theta_n - \theta_{n-1})^\top \nabla \mathcal{L}_{n-1}(\theta_{n-1})\\&+ \frac{1}{2} (\theta_n - \theta_{n-1})^\top \nabla^2 \mathcal{L}_{n-1}(\theta_{n-1})(\theta_n - \theta_{n-1}).
\end{aligned}
\end{equation}
By defining \(\Delta \theta \overset{\underset{\mathrm{def}}{}}{=} \theta_n - \theta_{n-1}\), Eq. (\ref{Eq:TaylorExpL1Theta2}) becomes:
\begin{equation}
    \begin{aligned}
        \mathcal{L}_{n-1}(\theta_n)& \approx \mathcal{L}_{n-1}(\theta_{n-1})+\Delta \theta^\top \nabla \mathcal{L}_{n-1}(\theta_{n-1})\\ & + \frac{1}{2} \Delta \theta^\top \nabla^2 \mathcal{L}_{n-1}(\theta_{n-1}) \Delta \theta. 
    \end{aligned}
\end{equation}
Here \(\nabla^2 \mathcal{L}_{n-1}(\theta_{n-1})\) is the Hessian matrix (a symmetric matrix of second-order partial derivatives, which is positive semi-definite)~\cite{vahedifar_2025_14631802}. Assuming that \(\theta_{n-1}\) is a local minimum, the model converges to a stationary point where the gradient vanishes, i.e., \(\nabla \mathcal{L}_{n-1}(\theta_{n-1}) \approx 0\), we can simplify the expression further:
\begin{equation}
       \mathcal{L}=\mathcal{L}_{n-1}(\theta_n) - \mathcal{L}_{n-1}(\theta_{n-1}) \approx \frac{1}{2} \Delta \theta^\top \nabla^2 \mathcal{L}_{n-1}(\theta_{n-1}) \Delta \theta.
\end{equation}
To derive an upper bound for \(\mathcal{L}\), the second-order term can be bounded using the maximum eigenvalue of the Hessian:
\begin{equation}
    \Delta \theta^\top \nabla^2 \mathcal{L}_{n-1}(\theta_{n-1}) \Delta \theta \leq \lambda_{\text{max}}^{n-1} \|\Delta \theta\|^2,
\end{equation}
where \(\lambda_{\text{max}}^{n-1}\) is the largest eigenvalue of \(\nabla^2 \mathcal{L}_{n-1}(\theta_{n-1}^*)\). Using these bounds, \(\mathcal{L}_{n-1}\) can be bounded by:
\begin{equation}
    \mathcal{L}_{n-1} \leq \frac{1}{2} \lambda_{\text{max}}^{n-1} \|\Delta \theta\|^2.\label{Eq: Upper bound}
\end{equation}
\section{Results \& Discussion}
\begin{table*}[t]
\centering
\caption{Inference Time ($ms$) Across Datasets and Parameters}
\label{tab: inference_times}
\resizebox{\textwidth}{!}{ 
\begin{tabular}{clccccccccc}
\cline{2-11}
&\textbf{Features}
 & {Position X} & {Position Y} & {Position Z} & {Velocity X} & {Velocity Y} & {Velocity Z} & {Force X} & {Force Y} & {Force Z}\\
\hline
\multirow{7}{*}{\rotatebox{90}{\textbf{Datasets}}}
&Drag Max Stiffness Y & 8.2 & 7.5 & 9.1 & 6.8 & 7.2 & 8.5 & 12.3 & 11.7 & 10.9 \\
&Horizontal Movement Fast & 6.5 & 7.1 & 6.9 & 7.4 & 6.2 & 7.8 & 9.5 & 10.2 & 8.7 \\
&Horizontal Movement Slow & 9.8 & 10.3 & 11.2 & 8.9 & 9.5 & 10.1 & 15.2 & 14.7 & 13.8 \\
&Tap and Hold Max Stiffness Z-Fast & 7.3 & 6.9 & 8.2 & 7.1 & 6.8 & 7.9 & 11.5 & 12.1 & 10.3 \\
&Tap and Hold Max Stiffness Z-Slow & 12.5 & 13.2 & 14.7 & 11.8 & 12.4 & 13.1 & 18.9 & 19.5 & 17.3 \\
&Tapping Max Stiffness Y-Z & 15.2 & 16.8 & 17.5 & 14.3 & 15.1 & 16.4 & 21.3 & 22.1 & 20.7 \\
&Tapping Max Stiffness Z & 14.7 & 15.3 & 16.8 & 13.9 & 14.6 & 15.2 & 20.5 & 21.3 & 19.8 \\
\hline
\end{tabular}%
}
\end{table*}
To the best of our knowledge, this is the first work to apply Leader-Follower learning from GT for signal prediction in the TI, demonstrating its effectiveness as a baseline method for remote robotic control.

\textbf{Datasets:} We utilized real-world haptic data traces for realistic data traffic in the experiments. The datasets capture kinaesthetic interactions recorded using a Novint Falcon haptic device within a Chai3D virtual environment. The datasets provide detailed records of 3D position, velocity, and force measurements. The datasets include five distinct types of movements, which are: 1) Tapping, 2) Tap and Hold, 3) Holding, 4) Free Air Movement (Horizontal Movement), and 5) Shearing (Diagonal Contact with Ground). A deadband mechanism was implemented to enhance perceptual relevance by filtering out minor velocity and force variations imperceptible to human users. This process optimizes data quality by reducing noise and improving relevance. The velocity and force perceptual deadband thresholds were set at 10\%~\cite{rodriguez_guevara_2025_14924062}.

\textbf{Neural Network:} We employed a 12-layer fully connected NN on the human operator's side and an 8-layer fully connected NN on the remote robot's side, with each layer containing 100 ReLU-activated units. The network weights were initialized using He initialization~\cite{He_2015_ICCV}. The selection of 12 and 8 layers likely reflects a balance optimized through experimentation. It is worth mentioning that increasing the robot's network complexity beyond eight layers provided diminishing returns in terms of prediction accuracy, while also risking increased latency.  

\textbf{Training:} Training was performed using stochastic gradient descent (SGD) with a momentum of 0.9, an initial learning rate of 0.01, and a batch size of 32.  While a higher learning rate accelerates weight updates, it also introduces instability as the objective function evolves, making careful tuning essential. Dropout~\cite{Dropout10.5555/2627435.2670313} was applied to train the discriminator net. The number of steps applied to the LeFo is a hyperparameter that we obtained by the cross-validation set. We trained all datasets with one NVIDIA RTX A6000 GPU.
\begin{figure}[t]
    \centering
    \includegraphics[width=\linewidth]{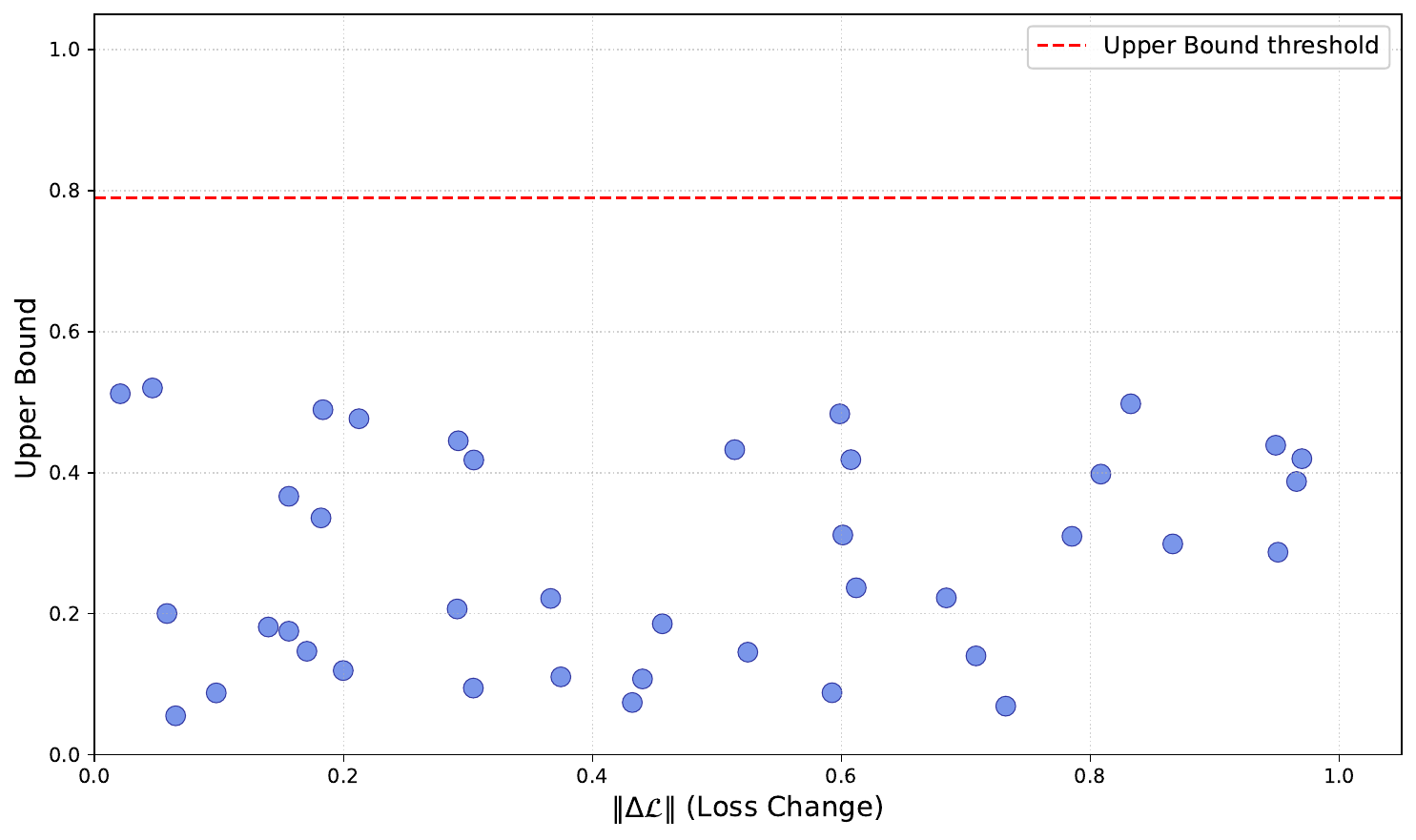}
    \caption{Empirical verification of Eq.~\ref{Eq: Upper bound}, for Drag Max Stiffness Y dataset.}
    \label{fig: Empirical verification}
\end{figure}
\begin{figure}[t]
    \centering
    \includegraphics[width=\linewidth]{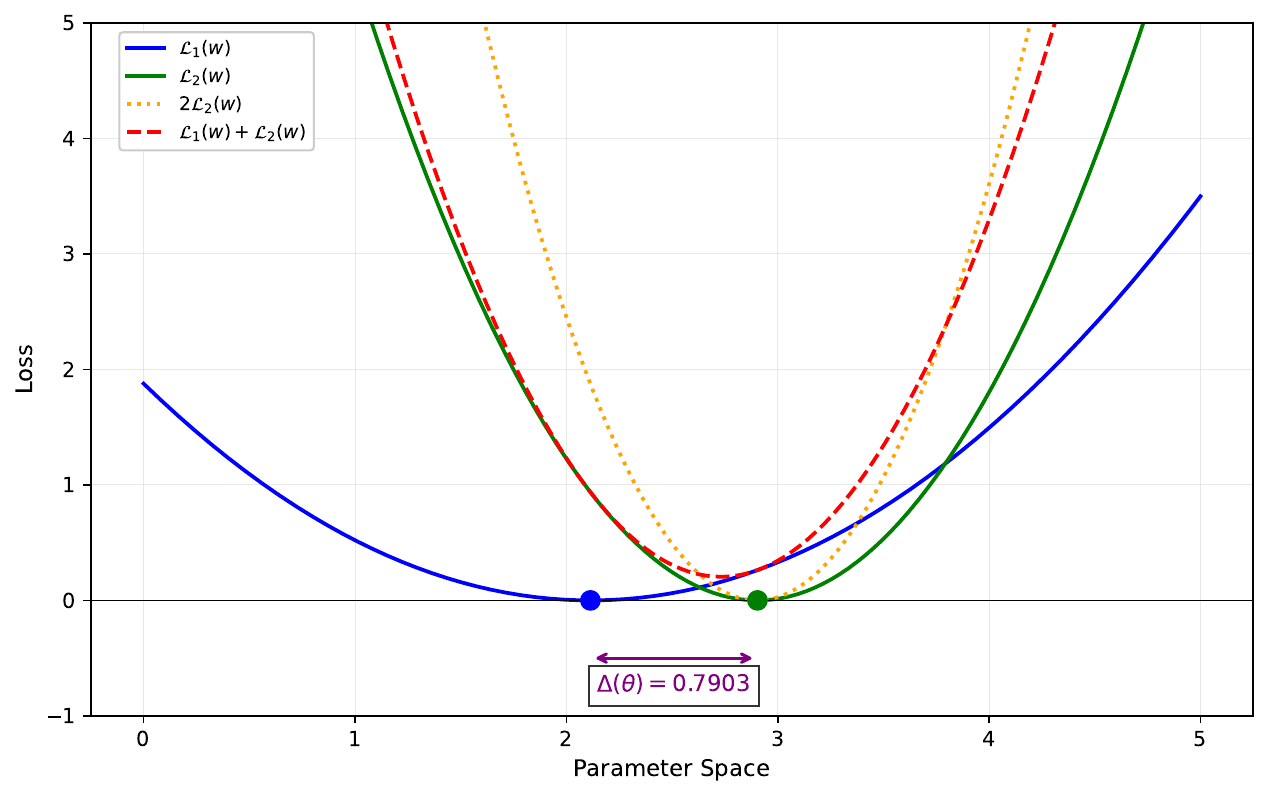}
    \caption{Parameter Space of Eq.~\ref{Eq: Upper bound}, for Drag Max Stiffness Y dataset.}
    \label{fig: Parameter Space}
\end{figure}

\textbf{Results:} Fig.~\ref{fig: Prediction accuracy} demonstrates the prediction accuracy of $\mathcal{H}$ and $\mathcal{R}$ across all datasets. Consistent with our theoretical expectations, $\mathcal{H}$ achieves superior accuracy, supporting the MiniMax game formulation where the Leader's role yields better performance. In the NN architecture, the human operator benefits from a deeper network to handle the complex task of predicting and generating high-dimensional haptic signals while accounting for latency constraints. In contrast, the robot operates effectively with a shallower network that prioritizes and minimizes computational overhead. 

Comparing features demonstrates that the force feature exhibits the most significant accuracy differences between Leader and Follower, often exceeding 8–12\%. At the same time, position metrics generally show smaller gaps, suggesting that they are more predictable or less sensitive to human-robot differences. Additionally, velocity metrics show better parity but still favor the Leader.

Table~\ref{tab: inference_times} presents inference time. While the current implementation does not meet the 1 $ms$ latency constraint in the teleoperation system, we expect methods for inference acceleration (e.g., a lightweight NN model) can be used with marginal accuracy loss. The LeFo method shows the feasibility of signal prediction in TI. The ``Horizontal Movement Fast" dataset is the most efficient, with average inference times ranging from 6.5 $ms$ to 10.2 $ms$. The ``Drag Max Stiffness Y" dataset also exhibits low latency. In contrast, datasets involving tapping yield the highest inference times, reaching up to 22.1 $ms$ for the force Y feature in the ``Tapping Max Stiffness Y-Z" dataset. These tasks likely require more computational effort due to the complexity of processing dual-axis stiffness. We posit that the variation in inference time across different feature sets may be attributed to multiple underlying factors. These include data complexity, sparsity, inter-feature dependencies, and correlations between agents, among others.

Additionally, we examine the relationship between loss curvature and the upper bound to assess the reliability of predictions. Fig.~\ref{fig: Empirical verification} empirically validates this relationship, as the loss in all the experiments is below the upper bound. The tightness of our upper bound is carefully justified, as higher-order Taylor expansion terms contribute negligibly to the loss. We maintain analytical tractability by retaining only the dominant first- and second-order terms, while ensuring the bound remains practically useful. Empirical results confirm that the omitted terms fall well below operational tolerances, validating our approximation approach. This yields a computationally efficient yet theoretically sound guarantee for maximum signal loss. Fig.~\ref{fig: Parameter Space} visualizes the parameter space through Hessian approximation using the largest eigenvalues. In Fig.~\ref{fig: Parameter Space}, the dots represent different NN training frameworks with different settings (e.g., with and without dropout, with
and without learning rate decay, different learning decay values, different initial learning rates, and different batch sizes).

\section{Conclusion}
In this work, we introduced the LeFo algorithm, a cooperative game framework designed to mitigate the impact of missing signals due to transmission error or packet loss in TI. By leveraging historical interaction data, LeFo enables adaptive prediction. Our formulation assigns distinct roles to the human side $\mathcal{H}$, which maximizes mutual information $I$, while the robot side $\mathcal{R}$ minimizes $\textit{KL}$ divergence, creating a resilient closed-loop system. The Taylor expansion establishes an upper bound for the loss function, ensuring guaranteed robustness against signal loss. Empirical results validate the effectiveness of the MiniMax game signal prediction in TI. Future work will explore feature selection to optimize the trade-off between accuracy and inference time.

{\small
\bibliographystyle{IEEEbib}
\bibliography{refs}
}
\end{document}